# A novel Deep Structure U-Net for Sea-Land Segmentation in Remote Sensing Images

Pourya Shamsolmoali, Masoumeh Zareapoor, Ruili Wang*, Huiyu Zhou, Jie Yang

*Abstract*--Sea-land segmentation is an important process for many key applications in remote sensing. Proper operative sea–land segmentation for remote sensing images remains a challenging issue due to complex and diverse transition between sea and lands. Although several Convolutional Neural Networks (CNNs) have been developed for sea-land segmentation, the performance of these CNNs is far from the expected target. This paper presents a novel deep neural network structure for pixel-wise sea-land segmentation, a Residual Dense U-Net (RDU-Net), in complex and high-density remote sensing images. RDU-Net is a combination of both downsampling and upsampling paths to achieve satisfactory results. In each down- and up-sampling path, in addition to the convolution layers, several densely connected residual network blocks are proposed to systematically aggregate multi-scale contextual information. Each dense network block contains multilevel convolution layers, short-range connections and an identity mapping connection which facilitates features re-use in the network and makes full use of the hierarchical features from the original images. These proposed blocks have a certain number of connections that are designed with shorter distance backpropagation between the layers and can significantly improve segmentation results whilst minimizing computational costs. We have performed extensive experiments on two real datasets Google-Earth and ISPRS and compare the proposed RDU-Net against several variations of Dense Networks. The experimental results show that RDU-Net outperforms the other state-of-the-art approaches on the sea-land segmentation tasks.

*Index Term*- Deep neural network (DNN), dense network (DenseNet), remote sensing images, sea-land segmentation, U-Net.

This research is partly supported by the National Science Foundation of China, (No: 61572315, 6151101179), the 973 Plan, China (No. 2015CB856004), the Marsden Fund of New Zealand, and H. Zhou was supported by UK EPSRC under Grant EP/N011074/1, Royal Society-Newton Advanced Fellowship under Grant NA160342, and European Union's Horizon 2020 research and innovation program under the Marie-Sklodowska-Curie grant agreement No 720325.

P. Shamsolmoali is with the Institute of Image Processing and Pattern Recognition, Shanghai Jiao Tong University, Shanghai 200240, China.

M. Zareapoor is with Institute of Image Processing and Pattern Recognition, Shanghai Jiao Tong University, Shanghai 200240, China.

R. Wang (corresponding author) is with the School of Logistics and Transportation, Central South University of Forestry and Technology, China, and the School of Natural and Computational Sciences, Massey University, Auckland, New Zealand (e-mail: ruili.wang@massey.ac.nz).

H. Zhou is with the Department of Informatics, University of Leicester, Leicester LE1 7RH, United Kingdom.

J. Yang is with the Institute of Image Processing and Pattern Recognition, Shanghai Jiao Tong University, Shanghai 200240, China (e-mail: jieyang@sjtu.edu.cn).

## I. INTRODUCTION

Machine vision is a technique of electronics field which widely applied in modern Remote Sensing Imagery. Remote sensing image segmentation, especially sea-land segmentation, has an important function in numerous fields such as coastline extraction [1] and maritime safety [2]. But up to now, feature extraction in remote sensing images, especially in a crowded scene, is a challenging task for sea-land segmentation. Continuous efforts have been made in the field. For instance, contrast to traditional thresholding segmentation models, Xia et al. [3] presented a model in which gray intensity features and local binary pattern features are combined. Ma et al. [4] and Liu et al. [33] presented a sea-land segmentation hierarchical model which reduced the computational costs. These approaches increased the segmentation accuracy; however, the models are not stable due to the compound intensity and texture distributions. There are some items like inland water, ships, and islands, which can confuse the algorithms and affect the segmentation results in high-resolution remote sensing images. Hence, the established classification models need to be improved.

Akbarizadeh [54] proposed a model called KWE, to extract texture features by using wavelet transform; that forms a feature vector composed of kurtosis values of wavelet energy features in SAR images and the model uses feature vectors and a level set function for the segmentation of textures. Tirandaz and Akbarizadeh [55] proposed a model named KCE that uses a single-stage curvelet to extract the texture feature and a new term is introduced based on the kurtosis feature value of the curvelet coefficients energy of the SAR image. At the last stage, a level set method is used to outline the boundaries between textures. [56] reported a level set method by using global and local information. In addition, a Gaussian convolution is applied to regularizing the level set function for the purpose of avoiding the computationally expensive re-initialization. In [57], the authors proposed a method for remote sensing image segmentation, which utilizes both spectral and texture information. Meanwhile, linear filters are used to provide enhanced spatial patterns. [58] took the space computing capacity of Cellular Automata and the data pattern search ability of Extreme Learning Machine based on Cellular Automata for edge detection in remote sensing images. Yao et al. [59] proposed a novel image segmentation method for remote sensing based on adaptive cluster ensemble learning. The clustering parameter of each image is calculated with affinity propagation automatically. Then, multiple clusters are trained separately and the predictions of them are combined under the ensemble learning framework. The authors in [60], based on the use of the co-association matrix and sub-clusters proposed computationally efficient methods of constructing ensembles of nonparametric clustering algorithms for satellite image segmentation. [61] Presents a multilayer perceptron neural network model to identify surface water in Landsat 8 satellite images. [62] Proposed a multilayer fusion model for adaptive segmentation and change detection of optical remote sensing image series. The method applies unsupervised or partly supervised clustering on a fused-image series by using cross-layer similarity measure, followed by multilayer Markov random field segmentation. Shen et al. [63] proposed a double-group Particle Swarm Optimization algorithm to improve the performance of the remote sensing image segmentation. The proposed model uses a double-group based evolution framework. [64] Used a classical Fuzzy C-means (FCM) method was for the coastline detection, but had been improved by combining the Wavelet decomposition algorithm to better suppress the inherent speckle noises of SAR image. In [65], [66] the authors proposed an end-to-end framework called multiple feature pyramid network (MFPN). In MFPN, an effective feature pyramid and a tailored pyramid pooling module are implemented that takes advantage of multilevel semantic features of high resolution remote sensing images.

In the past several years, deep learning delivers state-of-the-art performance for image segmentation [5], [9], classification [6] [45], tracking and detection [7] [8]. Since 2013, numerous Deep Convolutional Neural Network (DCNN) architectures have been designed and applied to various tasks such as, GoogleNet [10], Residual Net [11], and DenseNet [12]. Meanwhile, various models are used for semantic image segmentation. For example, Fully-connected Convolution Network (FCN) has demonstrated significant performance in image segmentation [2]. SegNet [13] is another model which used a FCN for image segmentation.

Liu et al. [14] proposed an algorithm for sea-land segmentation depending on the analysis of sea surface. Firstly, the sea surface of the corresponding input images is processed using the proposed model, and the sea sections are blocked out. Secondly, the statistical parameters of the sea region are assessed from the detected sea regions. In the end, to perform segmentation, a compatible thresholding according to the difference of the variances of the sea part and the other parts was applied. However, the overall accuracy decreases when texture features are similar at the coastline area [14], [18]. Li et al. [5] presented a model called DeepUNet which is deeper than U-Net and used DownBlocks for feature extraction and UpBlock for up-scaling. This model has good segmentation results compared to U-Net. Nonetheless, this model does not achieve good segmentation accuracy when the images have smooth sea–land boundaries or complex structures. To solve such problems, we intend to develop a new deep network architecture for end-to-end pixel-wise sea-land segmentation, named RDU-Net. RDU-Net uses convolution layers and multi-scale densely connected residual network blocks in each layer of the network. Its details are presented in Section 3.B.

The Residual Network (ReseNet) [11] and its modifications enable feature re-usage while the Dense Network (DenseNet) [12] enables new feature detection, both essential for representation learning. By conducting the combination of these two networks, the network shares mutual features while maintaining the flexibility to explore new features through double path architectures. In the proposed DenseNet block, we establish a certain number of connections in each feature layer and place them in the proper order. We intend to reduce the distance between the layers during the backpropagation that has a significant role in reducing the computation costs. Meanwhile, short-range connections and identity mapping [44] are combined to build a more effective network than the current approaches [5], [15], where identity mapping is a fundamental aspect of learning in deep networks and has significant effects in the training of the networks [24].

To improve the multilayer learning of RDU-Net, we add the densely connected residual network blocks at both down- and up-sampling paths. Compared to the standard DenseNets at depth $L$, the proposed DenseNet costs only $L \log L$, instead of $O(L^2)$ run-time complexity. Moreover, the proposed DenseNet slightly increases the short distance between the layers while the backpropagation increases from $1$ to $\log L+1$. Therefore, the proposed DenseNet can achieve promising results without increasing the GPU memory which is essential for ordinary DenseNets [43]. Therefore, the convolution layers can learn to collect more detailed outputs based on the provided features. Additionally, to significantly extract high-level features with minor loss errors, the proposed DenseNet blocks act as the connections between the multilevel convolution layers, in which the deep features will be concatenated before and after the convolution layers. This architecture helps us to bypass the redundant convolution procedure and deliver effective results. The main contributions of this work are summarized as follows:

- RDU-Net introduced a new DCNN architecture for remote-sensing sea-land segmentation. The proposed method is based on U-Net which contains densely connected residual network blocks that significantly improve the overall results while having low computation costs.
- To assess the performance of the proposed model, we perform a wide range of comparisons between U-Net [9], FusionNet [15], DeepUNet [5], and RDU-Net. The experiments show that RDU-Net is far more efficient than the other state-of-the-art networks.

## II. RELATED WORK

Lots of research efforts have been focused on remote sensing image segmentation in the last two decades. For an overall introduction, we refer the reader to textbook [16].

In the beginning, the researches mainly focused on the classical machine learning algorithms for handling the segmentation problems in remote sensing images, for example, Mohamed et al. [35] presented an automated method to select the parameters of the Gaussian Redial Basis Function kernel and used SVM regression to solve segmentation problems. Amini et al. [39] proposed an object classification model for hyperspectral data, based on a Random Forest algorithm.

Basaeed et al. [17] proposed a supervised hierarchical segmentation for remote sensing images. This model achieves segmentation via learning feature detection. The proposed model generates a group of convolutional layers to conduct multi-scale analysis on individual bands including different confidence maps on regional boundaries. Hu et al. [49] proposed a framework for evaluating and estimating the optimal scale parameter for the region-merging segmentation. In [51], the author presented a new technique for developing an oil spill segmentation, which effectively detects oil spill regions in blurry synthetic aperture radar images.

Limited works have been done on sea-land segmentation and coastline extraction based on colored imagery. Most of the present works used thresholding algorithms, complemented by morphological processes to reduce errors in the results. For instance, Ma et al. [4] proposed an algorithm by combining a modified Otsu's method with homogeneous textures and intensity features. Xia et al. [3] proposed a model, in which the original gray-level hyperspectral image is combined with the texture features extracted from local binary patterns to generate the combined feature map. Liu and Jezek [19] and Li et al. [20] presented thresholding algorithms to separate water areas from land areas. Zhang and Li [21] developed an algorithm based on the minimum class mean absolute deviation (MCMAD) for remote sensing image segmentation. Mao et al. [22] extended the standard Chan-Vese (CV) technique by two multifaceted wavelet transforms. Wang et al. [23] illustrated a novel supervised learning model for sea–land segmentation. Several features were firstly extracted from the entire image and later used to learn a robust sea–land classifier which converts the segment issues into a binary classification problem. In [24], the authors developed a technique to perform semantic segmentation for remote sensing images that uses a multi-scale model without increasing the number of parameters via optimization. The key idea is to train a dilated network with different patch sizes, to gain multi-scale features from heterogeneous contexts.

Silveira and Heleno [25] proposed a stylish technique by using area-based level sets and accepting a conglomerate of Log-normal solidities as the probabilistic architecture for the pixel intensities in the water and the land regions mutually. Chenglin et al. [26] presented a segmentation algorithm based on the moderate Chan-Vase technique. The authors presented a new architecture for sea/land segmentation in Synthetic Aperture Radar (SAR) images using level sets and a combination of Log-normal densities as the probabilistic model to describe water and land areas. Zhong et al. [27] created a method, based on the framework of geodesic remoteness. The method segments the sea-land as per both corresponding information of the spot and the land covers, which helps fast point-wised segmentation. These learning approaches depend on the physically designated features in a wide range. Consequently, for remote sensing imagery, while having composite information, such techniques have many misclassified pixels. For example, the green and shadow regions of the land are possibly classified as a single region. In addition, noise in water areas and waves could be treated as land regions.

Currently, to figure out a solution for such problems, we are motivated to use deep learning, due to its impressive performance in handling different computer vision tasks. The end-to-end Fully Convolutional Networks (FCNs), which was firstly developed by Long et al. [28], It initially performs pixel-wise segmentation by substituting FCN layers with CNN layers. Nonetheless, the FCNs create coarse segmentation maps due to the loss of data in subsampling actions. Therefore, researchers focused on delivering the pixel-wise segmentation models. In [44], the authors proposed the deep residual learning framework that uses new connections to simplify training, instead of using skip connection in deep networks. Lguensat et al. [29] proposed a model called EddyNet, a deep learning architecture for presetting eddy recognition and classification using the Sea Surface level maps supplied by the Copernicus Marine and Situation Monitoring Service. EddyNet contains a convolutional encoder-decoder layer for the pixel-wise classification.

There are some works to create links between the pooling and un-pooling layers. In the CNN, the max-pooling action is non-invertible; but, the rough inverse can be achieved by recording the positions of the maxima inside each pooling district with a set of adjustment variables. U-Net [9] is another effective FCN structure for image segmentation which is in fact used for processing biomedical images. Its structure contains a down-sampling path and an up-sampling path and its feature maps from the down-sampling path are collected and farther used for consistent up-sampling in the expansive path. DeepUnet [5] explores a new structure for pixel-wise water-land segmentation. The authors proposed to concatenate layers in the contracting path to create an expansive path. Therefore, sequential convolution layers can learn to collect more accurate outputs based on the extracted data. In our proposed model, in each down- and up-sampling path in addition to the convolution layers, there are several blocks in the proposed DenseNet, which contains multilevel convolution layer and combined connections, are implemented which are illustrated in the next section. Table I summarizes key deep learning achievements for image segmentation, where we listed the approaches, types of image data, tasks, computational complexity, and efficiency.

III. PROPOSED METHOD

In this section, we present the detailed structure of the proposed model.

*A. Overall Architecture*

Fig. 1 is a graphical description of the proposed RDU-Net architecture. The proposed network is built based on the U-Net structure [9]. U-Net is one of the popularly used models for image segmentation and already has achieved superior segmentation performance on several types of images and datasets [5], [36], [37] and these are the reasons why we choose the U-Net as the baseline. Moreover, the U-Net consists of a contracting path to extract appropriate features and an expansive path which supports the prediction of the synthesis.

TABLE I
SUMMARY OF SOME KEY DEEP LEARNING MODELS FOR OBJECT SEGMENTATION.

| Ref. | Approach | Data and Tasks | Optimization | Efficiency |
|---|---|---|---|---|
| [5] | Convolution blocks instead of convolution layers | Sea-Land Segmentation | U-connection and Plus connection | --- |
| [9] | Down sampling, up sampling and augmentation strategy for feature extraction | Biomedical Image Segmentation | --- | --- |
| [13] | Optimizing VGG by removing fully connected layers | Multi-class Image Segmentation | non-linear upsampling | --- |
| [15], [40] | Convolution and residual network | Biomedical Image Segmentation | Summation-based skip connections | --- |
| [17] | Multi-scale convolutional neural networks | Remote sensing image segmentation | Using fused confidence map | --- |
| [29] | Encoder-decoder and pixel-wise classification | Sea surface height | --- | --- |
| [37] | Residual U-Net | Remote sensing road extraction | Skip connections for information propagation | --- |
| [24] | Dilated convolution with distinct patch sizes, | Remote sensing segmentation | Distinct patch sizes | --- |
| This work | Identity mapping and residual DensNet in U-Net | Remote sensing segmentation | Layers distance reduction while backpropagation | Dense Growth rate gradual raising |

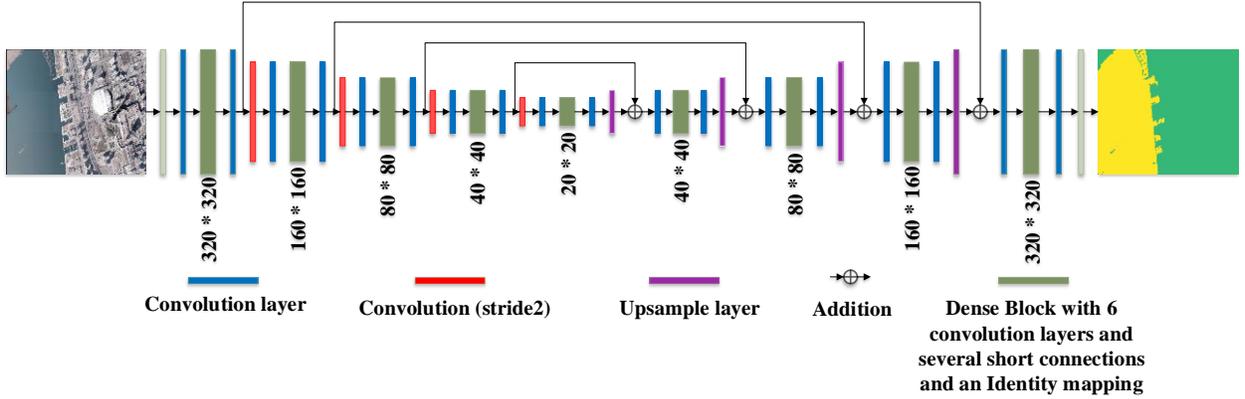

Fig. 1. The architecture of RDU-Net. The contracting path (top to center) and the expansive path (center to end). Each densnet block has several identity mapping connections in the same path, but the long identity mapping connections join two different paths.

Four kinds of structure blocks are used to build RDU-Net. Each blue block is a general convolutional layer. There are various combinations for the convolution layers, batch normalization (BN) and activation (PReLU) [41] to reduce the error rate and computational costs. He et al. [44] discussed the effects of different combinations and proposed a full pre-activation design. In this work, we also used a full pre-activation model in RDU-Net. The PReLU activation function is treated base-wise. If $F_{i-1}$ is the input, the outcome of the $i^{th}$ convolution layer is calculated as:

$$F_i = max\ (0,\ w_i \sim F_{i-1} + b_i) \quad (1)$$

where $w_i$ are the weights; $b_i$ are the biases in the layers and $\sim$ means either convolution or deconvolution action. In the proposed model, Adamax [30] is used to derive the best weights and biases. Each green block is a proposed DenseNet block. The red blocks are convolutions with stride 2 instead of max-pooling to perform down-sampling for feature compression. The purple blocks are the up-sampling layers in the expansive path to up-sample the data instead of deconvolution blocks in the U-Net. Table II shows the details of the proposed network.

Each level in the expansive path commences with an up-sampling layer. These layers un-pool the features to a larger level, and then they merge the features with the feature map of the equal level at the contracting path via an extensive identity mapping connection [44], which is further explained below.

The original residual unit mainly acts as follows:

$$y_l = h(x_l) + F(x_l, W_l), \quad (2)$$
$$x_{l+1} = f(y_l). \quad (3)$$

while $x_l$ is the input feature to the $l^{th}$ residual unit. $W_l = \{w_{l,k|1 \leq k \leq K}\}$ is a set of weights and biases are associated to the $l^{th}$ residual unit, and $K$ is the layers number in a residual unit. $F$ represents the residual function (the number of layers and kernel size). After the element-wise addition, function $f$ is performed which is ReLU in [11] but in RDU-Net, PReLU is applied. The function $h$ is the identity mapping: $h(x_l) = x_l$.
If $f$ is identity mapping: $x_{i+1} \equiv y_i$, by combining Eqs. (2) and (3), we can have

$$x_{l+1} = x_l + F(x_l, W_l). \quad (4)$$

Recursively, we get $(x_{l+2} = x_{l+1} + F(x_{l+1}, w_{l+1}) = x_l + F(x_l + w_l) + F(x_{l+1}, w_{l+1}), etc.)$, and

$$x_L = x_l + \sum_{i=l}^{L-1} F(x_l, w_l), \quad (5)$$

Eq. (5) shows some interesting assets for both deeper $L$ and shallower $l$ units. Firstly, in addition to the residual function, which is in the form of $\sum_{i=l}^{L-1} F$, the feature $x_L$ of the deeper unit $L$ can be denoted as feature $x_l$ of any shallower unit $l$, that presents the model in a residual style. Secondly, the feature $x_L = x_0 + \sum_{i=0}^{L-1} F(x_i, w_i)$, of any deep unit $L$, is the outline of the results of all the earlier residual functions (plus $x_0$). Moreover, Eq. (5) leads to proper backpropagation properties. By representing the loss function as $\varepsilon$, from the main instruction of backpropagation [44], we observe:

$$\frac{\partial \varepsilon}{\partial x_l} = \frac{\partial \varepsilon}{\partial x_L} \frac{\partial x_L}{\partial x_l} = \frac{\partial \varepsilon}{\partial x_L}\left(1 + \frac{\partial}{\partial x_l}\sum_{i=l}^{L-1} F(x_i, w_i)\right). \quad (6)$$

Eq. (6) shows that the gradient $\frac{\partial \varepsilon}{\partial x_l}$ can be divided into two parts: a part of $\frac{\partial \varepsilon}{\partial x_L}$ that directly propagates information without using the weight layers, and the other part is $\frac{\partial}{\partial x_l}\sum_{i=l}^{L-1} F$ that uses the weight layers for propagation.

By using the above equations and their adjustments, we could split the identity mapping (shortcut), $h(x_l) = \lambda_l x_l$:

$$x_{l+1} = \lambda_l x_l + F(x_l, w_l), \quad (7)$$

while $\lambda_l$ is a controlling scalar. Recursively, by using this formulation, an equation similar to Eq. (5), can be obtained:

$$x_L = \left(\prod_{i=l}^{L-1} \lambda_i\right) x_l + \sum_{i=l}^{l-1}\left(\prod_{j=i+1}^{L-1} \lambda_j\right) F(x_i, w_i), \quad (8)$$

Eq. (8) can be simplified as:

$$x_L = \left(\prod_{i=l}^{L-1} \lambda_i\right) x_l + \sum_{i=l}^{L-1} \hat{F}(x_i, w_i), \quad (9)$$

Similar to Eq. (6), the backpropagation function can be formulated as follows:

$$\frac{\partial \varepsilon}{\partial x_l} = \frac{\partial \varepsilon}{\partial x_L}\left(\left(\prod_{i=l}^{L-1} \lambda_i\right) + \frac{\partial \varepsilon}{\partial x_l}\sum_{i=l}^{L-1} \hat{F}(x_i, w_i)\right). \quad (10)$$

In Eq. (9), the first additive term is controlled by a factor $\prod_{i=l}^{L-1} \lambda_i$. In the very deep networks, if $\lambda_i > 1$ in all $i$, the factor can be exponentially large; if $\lambda_i < 1$ in all $i$, the factor should be exponentially small and vanish, that stops the backpropagated signals passing through the shortcut and run via the weight layers.

Additionally, in RDU-Net, at the contracting path, after having performed every down sampling, the sum of feature maps is doubled. Once passing the contracting path, the bridge layer (20×20), a proposed DenseNet block, starts to increase the feature maps in the next expansive level. In the expansive part, the number of the feature maps is split in every level to keep the balance of the network. Before and after each DenseNet block, there are two convolutional layers. These convolutional layers perform as a connector to pass the input feature maps to the DenseNet blocks since the feature maps from the earlier layer may be different from the DenseNet block. An additional advantage of these convolutional layers on both sides of the dense block is to keep the entire network balanced, as shown on the left side of Fig. 2.

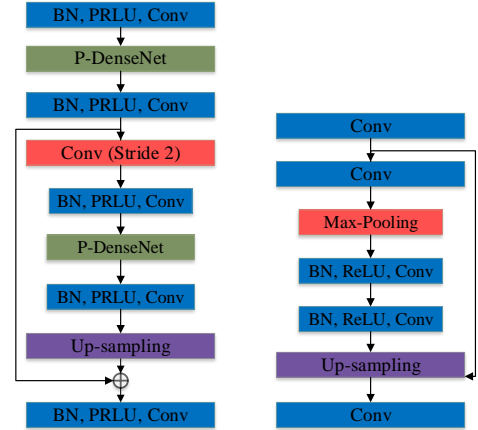

Fig. 2. The core difference between U-Net (right) and RDU-Net (left). RDU-Net is not only much deeper in comparison with U-Net but also has proposed residual denseNet blocks (P-DenseNet) and several short-range connections and identity mapping which improves the overall performance with less generalization errors and better computation efficiency.

One of the concepts in CNNs is the use of the receptive field of a unit in a certain layer in the network. Since an area in an input image outside the receptive field of a unit does not affect the value of that unit, it is necessary to cautiously control the receptive field to ensure that it covers the entire relevant regions. In image segmentation where we make a prediction for each pixel in the image, it is critical for each output pixel to have a big receptive field, hence no important information is left out when making the prediction [53]. The receptive field size of a unit can be enlarged in several ways. One option is to stack more layers to make the network deeper. Another way to increase the receptive field size is sub-sampling. However, a very large receptive field can introduce noise to the network. In RDU-Net, to systematically increase the size of the receptive field, we adopt the model proposed in [47].

*B. Densely Connected Residual Network*

To start, we briefly present the difference between ResNet [11] and DenseNet [12], [32]. The residual path reuses features implicitly, but it does not have any impact on the exploration of new features. On the other hand, the densely connected network keeps extracting new features but suffers from certain redundancy [46]. In this section, we give details of our proposed architecture.

TABLE II
THE NETWORK DETAILS OF RDU-NET

| Block type | Ingredients | Kernel size | Size of feature maps |
|---|---|---|---|
| Input | | | 320×320×1 |
| Down 1 | Conv | 3×3 | 320×320×64 |
| | Dense (6 Conv) | 3×3 | -- |
| | BN, PReLU, Conv | 2×2 | 320×320×64 |
| | Conv (stride 2) | 2×2 | 160×160×64 |
| Down 2 | BN, PReLU, Conv | 3×3 | 160×160×128 |
| | Dense(6 Conv) | 3×3 | -- |
| | BN, PReLU, Conv | 2×2 | 160×160×128 |
| | Conv (stride 2) | 2×2 | 80×80×128 |
| Down 3 | BN, PReLU, Conv | 3×3 | 80×80×256 |
| | Dense(6 Conv) | 3×3 | -- |
| | BN, PReLU, Conv | 2×2 | 80×80×256 |
| | Conv (stride 2) | 2×2 | 40×40×256 |
| Down 4 | BN, PReLU, Conv | 3×3 | 40×40×512 |
| | Dense(6 Conv) | 3×3 | -- |
| | BN, PReLU, Conv | 2×2 | 40×40×512 |
| | Conv (stride 2) | 2×2 | 20×20×512 |
| Bridge | BN, PReLU, Conv | 2×2 | 20×20×1024 |
| | Dense(6 Conv) | 3×3 | -- |
| | BN, PReLU, Conv | 2×2 | 20×20×1024 |
| Up 4 | Unpooling | -- | 40×40 |
| | Addition | -- | -- |
| | BN, PReLU, Conv | 3×3 | 40×40×512 |
| | Dense(6 Conv) | 3×3 | -- |
| | BN, PReLU, Conv | 2×2 | 40×40×512 |
| Up 3 | Unpooling | -- | 80×80 |
| | Addition | -- | -- |
| | BN, PReLU, Conv | 3×3 | 80×80×256 |
| | Dense(6 Conv) | 3×3 | -- |
| | BN, PReLU, Conv | 2×2 | 80×80×256 |
| Up 2 | Unpooling | -- | 160×160 |
| | Addition | -- | -- |
| | BN, PReLU, Conv | 3×3 | 160×160×128 |
| | Dense(6 Conv) | 3×3 | -- |
| | BN, PReLU, Conv | 2×2 | 160×160×128 |
| Up 1 | Unpooling | -- | 320×320 |
| | Addition | -- | -- |
| | BN, PReLU, Conv | 3×3 | 320×320×64 |
| | Dense(6 Conv) | 3×3 | -- |
| | BN, PReLU, Conv | 2×2 | 320×320×64 |
| output | Conv | 1×1 | 320×320×1 |

Based on the analysis, we propose a gradual DenseNet architecture which shares $f_t^k(.)$ among the overall layers to keep the advantages of reusing features with low redundancy, and keep the densely connected path to build a network which has wide flexibility in learning new features. We formulate the proposed architecture as follows:

$$x^k \triangleq \sum_{t=1}^{k-1} f_t^k(h^t), \qquad (11)$$

$$y^k \triangleq \sum_{t=1}^{k-1} v_t(h^t) = y^{k-1} + \emptyset^{k-1}(y^{k-1}) \qquad (12)$$

$$r^k \triangleq x^k + y^k, \qquad (13)$$

$$h^k \triangleq g^k + r^k, \qquad (14)$$

where at the $k^{th}$ step in each distinct path, $x^k$ and $y^k$ represent the extracted information; $v_{t(.)}$ and $f_t^k(.)$ are the feature learning functions. Eq. (11) denotes the densely connected path which allows new features to be explored. Eq. (12) denotes the residual path that enables the re-use of the shared features, and Eq. (13) describes the double path that aggregates and passes them to the final transformation function shown in Eq. (14). The final transformation function $g^k(.)$ demonstrates the current state, which is used for further mapping or prediction. This supports the data flow over the network; and consequently handles the vanishing-gradient issue. Fig. 3 shows the proposed double path architecture that is being used in our experiments. Since the DenseNet is more used than the ReseNet in practice, we select the DenseNet as the backbone and add the identity mapping path to build the double path network.

In a single DenseNet block in the RDU-Net, there are 6 convolution layers with different kernel sizes. The first and the last one is $1\times1$ and the rest have a $3\times3$ kernel size. Additionally, there are several short-range connections and identity mapping [44] to add the input and the output of each unit. The output of each convolution layer, as the input to the next convolution layer, is combined with the output of the other convolution layers which are sequentially fed to the other convolution layers and a Batch normalization layer is added at the end of the last convolution layer. The Batch normalization layer has the ability to handle the interior covariate shift issues and speeds up the training process as it pushes the mean activation of the input data near 0 and the standard deviation near 1 [38].

$$BN(x) = \max(\frac{x - M(x)}{\sqrt{var(x)+e}} \cdot g + d, 0), \qquad (15)$$

where $x$ is the input; $M(x)$ and $Var(x)$ denote the mean and variance of the input; $\varepsilon$ is a minor constant value; $\gamma$ is the scalar, and $\delta$ is the shift parameter of a data batch.

The optimized DenseNet block can be represented as

$$F_{main}(x) = PReLU(F(x) + x), \qquad (16)$$

where

$$F(x) = BN(Conv(Conv \ldots (Conv(x, W_1), W_2), \ldots, W_6)). \qquad (17)$$

The standard DenseNet model has $p$ new feature maps from each layer, where $p$ is a constant stated as the *growth rate*. DenseNet tends to rely on high-level features more than low-level features. We implement the gradual increase of the growth rate as the depth grows. This increases the amount of the features coming from the following layers relative to those from the previous layers. The growth rate is set to $p = 2^{m-1}p_0$, while $m$ is the convolution layer index, and $p_0$ is a constant. Such growth rate adjustments do not generate any additional hyper-parameters. The strategy of *growth rate gradual increasing* creates a larger amount of parameters in the final layers of the model. This considerably improves the computational efficiency.

To improve the computational efficiency, we also adopt a model where with the same number of the connections, at the time of backpropagation, the distance between any two layers should be as short as possible [43]. In the proposed DenseNet, each $x_i$ is considered as a node in a graph, and the fixed edge ($x_i$, $x_j$) exists if $x_i$ directly receives inputs from $x_j$. The backpropagation distance (BD) between $x_i$ and $x_j$ $(i > j)$ is considered as the shortest path length from $x_i$ to $x_j$ on the graph. Additionally, we determine the maximum backpropagation distance (MBD) as the utmost BD between all the pairs $i > j$.

In this way, the MBD of DenseNet is 1, if we ignore the transition layers. To minimize the $O(L^2)$ computation cost of DenseNet, we used Log-DenseNet [43] which slightly increases MBD to $1 + log_2 L$ and only uses $O(L \log L)$ connections.

In the proposed DenseNet, each layer $i$ directly receives inputs from at most $\log(i) + 1$ of the prior layers, and these input layers are separated from depth $i$ with base 2.

$$x_i = f_i(concat(\{x_{i-[2^k]}: k = 0, ..., [\log(i)]\}); \theta_i) \quad (18)$$

For instance, the entry features for the $i^{th}$ layer are layer $i-1, i-2, i-4, ...$ We set the input index at layer $i$ to be $\{i - [2^k]: k = 0, ..., \lfloor\log(i)\rfloor\}$. Subsequently, the density of layer $i$ is $\log(i) + 1$, and the global complexity of the proposed DenseNet is $\sum_{i=1}^{L}(\log(i) + 1) \leq L + L \log L = \Theta(L \log L)$, which is considerably smaller than the complexity of the original DenseNet [12].

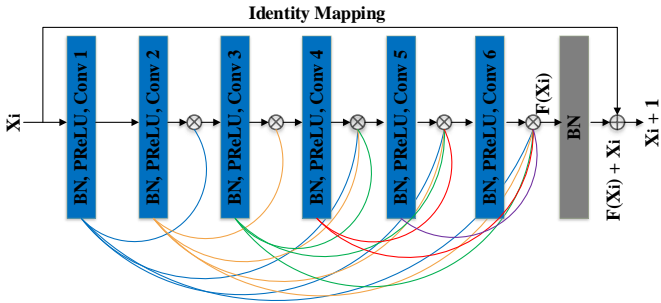

Fig. 3. The architecture of the proposed Residual DenseNet block. "Conv" denotes a convolutional layer, and "BN" represents a batch normalization layer $\otimes$ represents concatenation and $\oplus$ denotes addition.

Recently, ResNet models [40] show that numerous numbers of layers have limited contribution and can be randomly dropped while training. This convert the operation of ResNets like recurrent neural networks (RNN), but the numbers of the parameters in ResNets are significantly larger as each layer has its own weights. The proposed DenseNet obviously differentiates the information that is added to the network from the preserved information. DenseNet layers are very narrow (8 feature-maps per layer), adding only a small set of feature-maps to the network and keep the remained feature-maps intact. Hence, the last layer (classifier) makes the decision based on all the feature-maps in the network. In addition to its better parameter efficiency, one big advantage of DenseNets is that there are flows of information and gradients all over the network, which resulted in easy training. Each layer has short and direct accesses to the gradients from the loss function and the input signal, which bring an implicit deep supervision. As a direct result of the input concatenation, the feature maps learned by any of the DenseNet layers are accessible to all the subsequent layers. This stimulates feature reuse throughout the network, and leads to more well-set models. Further, dense connections show regularizing effects, which reduces over-fitting on tasks with smaller training sets.

As shown in Fig. 1, there are several short-range connections and identity mapping in the network to add the low-level and the high-level features. Hence, the combined feature maps are later fed as the input to the upscaling layers. Moreover, for the proper reconstruction, a single identity mapping [44] in the dense block has been used to merge the feature maps which are generated from the convolution layers. Hence, these novel components significantly improved the overall segmentation results and robustness of the RDU-Net as compared to U-Net [9], DeepUNet [5], and FusionNet [15].

IV. EXPERIMENTS DETAILS AND ANALYSIS

In the following section, we explain the details of the used datasets, experimental results, and compare the performance of the proposed model with the other models.

*A. Data augmentation and preprocessing*

In this paper, to evaluate the efficiency of RDU-Net, the images from Google Earth and ISPRS_Benchmark have been used. To prepare the dataset, we adjust the eye altitude of Google Earth to get the spatial resolution of around 3.5 meter per pixel and then use "GET Screen" to capture images. 396 large images from several geographical places are obtained. We crop all the large images manually and selected 1648 illustrative sea–land images with the size of 1500×1500 pixels.

Secondly, we randomly select 1592 images for the training set, 11 images for validation and 45 images for testing. Meanwhile, as the cropped images are unlabeled, we use Labelbox [42] which is a web based annotation tool to perform the labeling. Fig. 4 shows a few samples of the cropped sea–land images. To increase the productivity of the training, we select the images which cover both sea and land. Data augmentation is vital to show the desired robustness of the network, as we have insufficient training samples. For the implementations of RDU-Net, we have the following alterations for data augmentation [50].
(1) Random vertically and horizontally flipping.
(2) Random conversion by [−8, 8] pixels.
(3) Random scaling in the range [1, 1.5].

*B. Network Setup and Experiments*

We implement RDU-Net by using Keras 2.1.2, the deep learning open-source library [31] and TensorFlow 1.3.0 GPU as the backend deep learning engine. Python 3.6 is used for all the implementations. All the implementations of the network are conducted on a workstation equipped with an Intel i7-6850K CPU, a 64 GB Ram and an NVIDIA GTX Geforce 1080 Ti GPU and the operating system is Ubuntu 16.04.

As the data augmentation is used to duplicate the input images by having mirroring transformations and rotation for training and preparing the final results. The Softmax function has been used to classify the results. For the sea-land segmentation, RDU-Net sorts the pixels into sea, land or ship. Hence the Softmax function is used as follows. Assume that N different values should be acquired after classification, hence for the given input $x^{(i)}$, the possibility of its classification outcome $y^{(i)}$ which has 3 class is as follows:

$$p(y^{(i)} = 3|x^{(i)}, \theta) = \frac{e^{\theta_2^T x^{(i)}}}{\sum_{l=1}^{N} e^{\theta_l^T x^{(i)}}} \quad (19)$$

where θ is denoted as the model parameter, and T is transpose. The total probabilities of all the classes are one. The generated ground-truthed cost function and the classification results are shown as follows:

$$J(\theta) = -\frac{1}{m \times n}\left[\sum_{i=1}^{m}\sum_{j=1}^{n}\sum_{q=1}^{2} 1\{y^{(i,j)}=q\}\log(p_2^{(i,j)})\right] + \frac{\lambda}{2}\|\theta\|^2 \quad (20)$$

where $m$ and $n$ show the numbers of rows and columns; $\theta$ is the network parameter, and $(x^{(i,j)}, y^{(i,j)})$ and $P_2^{(i,j)}$ denote the ground truth of any pixel on row $i$ and column $j$ and the possibility to this pixel being classified as either sea or land. $1\{y^{(i,j)} = q\}$ is the stating function, where, if the ground truth of the consistent pixel is $q$, then one is obtained; otherwise, zero is obtained. For the regularization task, $\frac{\lambda}{2}\|\theta\|^2$ is used.

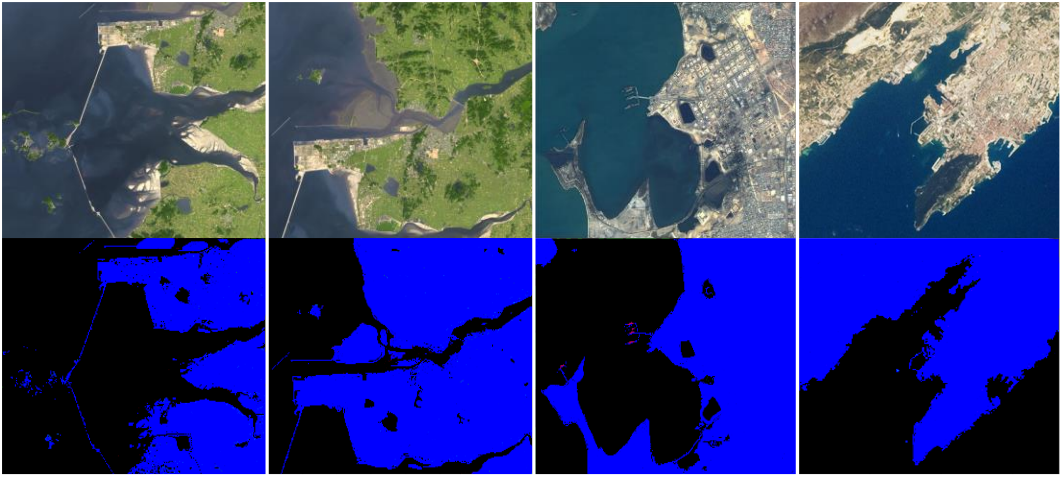

Fig. 4. Samples of collected sea–land images and the grand truth.

To evaluate the efficiency of RDU-Net, we compare it with the U-Net [9], FusionNet [15] and the DeepUNet [5] with the same datasets and experimental settings. The network structures of the mentioned models are either downloaded directly from the GitHub web pages or implemented with the technical details that the authors provided.

### C. Evaluation Metrics

The proposed model starts the training with the mini-batch size of 16. Initially, the learning rate is set to 0.001, and to guarantee a good learning result, the learning rate is divided by ten every fifteen epochs. There are 300 epochs during the training. The Adamax has been used as the optimizer to optimize the network for adjusting the parameters such as weights and biases.

$F_1$ score and the global pixel-wise accuracy of sea and land are used to evaluate the quantitative outcomes. $F_1$ score is representing the harmonic mean of precision and recall. They are calculated as follows:

$$F_1^i = 2 \times \frac{precision_i \times recall_i}{precision_i + recall_i} \quad (21)$$

while

$$precision_i = \frac{TP_i}{TP_i + FP_i}, \quad recall_i = \frac{TP_i}{TP_i + FN_i} \quad (22)$$

where $TP_i$ is the number of true positives for the classes; $FP_i$ and $FN_i$ represent the false positive and false negative, respectively. All these metrics are calculated by means of the pixel-based confusion matrices. Meanwhile, the overall accuracy can be calculated by normalizing the trace of the confusion matrix [5].

### D. Performance and Comparison

In the experiments, the quantitative analysis of the segmentation results of U-Net [9], FusionNet [15], DeepUNet [5] and our proposed method RDU-Net has been conducted. The used images contain both the scenes of harbors and islands and include a complicated distribution of texture and intensity. Some of the results are shown in Figs. 5, 6, 7, 8 and 9. In all these figures, we can clearly see that RDU-Net has significant performance in comparison with the other methods. Fig. 5 (a) represents the input image which contains a portion of the island. The land portion of the image has uneven surface colors. Fig. 5 (f) shows the result of RDU-Net. Compared to Fig. 5(c) U-Net, Fig. 5(d) FusionNet and Fig. 5(e) DeepUNet, the proposed model properly segments the sea and land portions without any errors. Table III illustrates the details of Fig. 5. We can see that RDU-Net's precision 2.2% is higher than U-Net, 1.16% higher than FusionNet and 0.66% higher than DeepUNet. The accuracy of RDU-Net is 3.33% higher than U-Net [9], 2.5% higher than FusionNet [15] and 0.63% higher than DeepUnet [5]. For the other parameters like recall and $F_1$ measure the RDU-Net also has better performance as compared to other state-of-the-arts models.

TABLE III
FIG. 5 EVALUATION RESULTS

| Models | Precision | Recall | $F_1$ | Accuracy |
|---|---|---|---|---|
| U-Net | 0.9734 | 0.9715 | 0.9532 | 0.9642 |
| FusionNet | 0.9852 | 0.9833 | 0.9711 | 0.9726 |
| DeepUNet | 0.9902 | 0.9859 | 0.9830 | 0.9912 |
| RDU-Net | **0.9968** | **0.9916** | **0.9897** | **0.9975** |

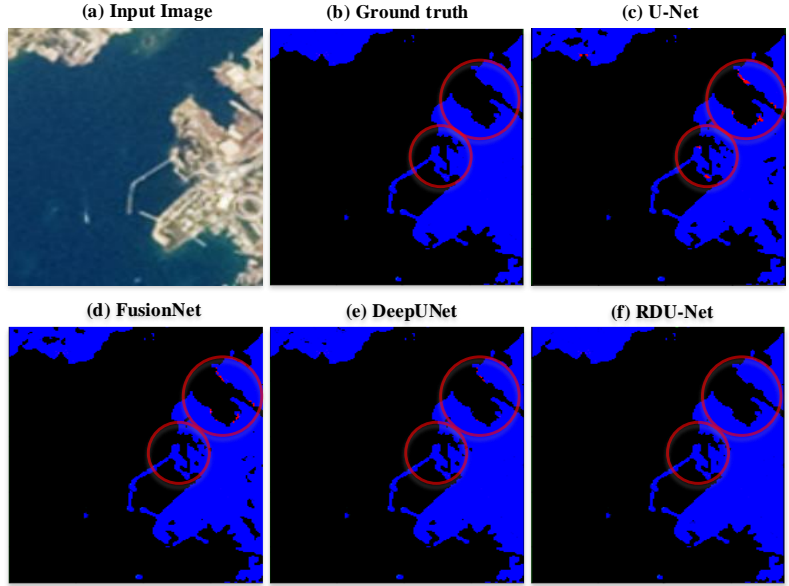

Fig. 5. The segmentation results from U-Net, FusionNet, DeepUNet, and RDU-Net.

Fig. 6 represents the segmentation results of different models. The tested image is more complicated than the tested image shown in Fig. 5. It contains small cars, ships, and shadows. Table IV lists the evaluation results of different models in Fig. 6. The results indicate that RDU-Net precision 3.51% is higher than U-Net, 2.239% higher than FusionNet [15] and 1.16% higher than DeepUnet [5]. The overall accuracy of RDU-Net is 6.51% better than U-Net, 5.44% higher than FusionNet [15] and 2.76% higher than DeepUNet [5]. The $F_1$ result of RDU-Net is 2.73% higher than DeepUnet, 4.77% higher than FusionNet and 5.87% higher than U-Net.

As previously discussed, the main reason for the higher performance of RDU-Net, in comparison with other state-of-the-art models, is firstly due to densely connected residual blocks which are equipped with shorter distance backpropagation that allows deep features to be extracted from the input images, and the other reason is the usage of several identity mapping connections all over the network which provided feature reuse for RDU-Net.

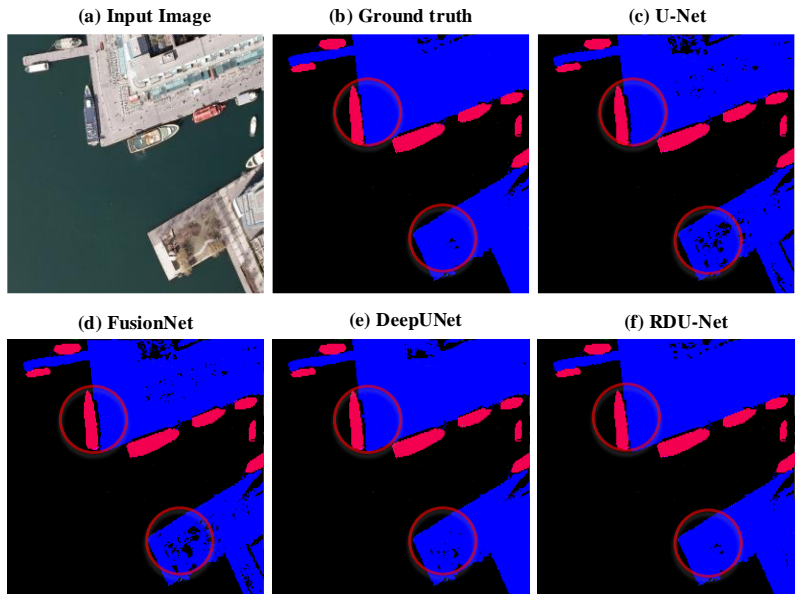

Fig. 6. The segmentation result of the high-resolution complex area by different models.

TABLE IV
FIG. 6 DETAILS EVALUATION RESULTS

| Models | Precision | Recall | $F_1$ | Accuracy |
|---|---|---|---|---|
| U-Net | 0.9047 | 0.9033 | 0.9026 | 0.9056 |
| FusionNet | 0.9125 | 0.9089 | 0.9079 | 0.9192 |
| DeepUNet | 0.9311 | 0.9371 | 0.9283 | 0.9232 |
| RDU-Net | **0.9593** | **0.9615** | **0.9437** | **0.9658** |

Figs. 7 and 8 show the results of different models while the input images are very complicated. The input images contain very small ships which stand very close to each other, the green spaces, shadows and tiny sea-land boundary. Generally, such factors significantly affect the segmentation result and make the operation task difficult. As it is visible in Figs. 7 and 8 where all the small ships and minor objects are properly detected and segmented out from the sea area by the proposed RDU-Net. The results of U-Net [9] and FusionNet [15] are not satisfactory because the input images have high-resolution, high density and contain objects of various scales. The experiments prove that the U-Net [9] and FusionNet [15] are not able to deal well with the complex areas. DeepUNet [5] has better performance as compared to the other two models. Still, it has some missing segmentation at the boundaries and shadow areas. Tables V and VI are respectively showing the performance of Figs. 7 and 8. The best result is bolded in each table.

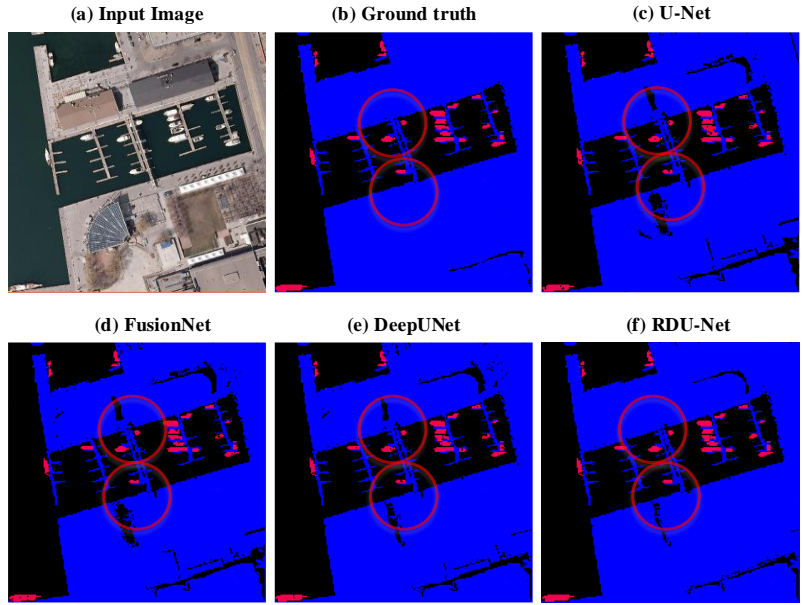

Fig. 7. The segmentation result of the high-resolution complex area by different models.

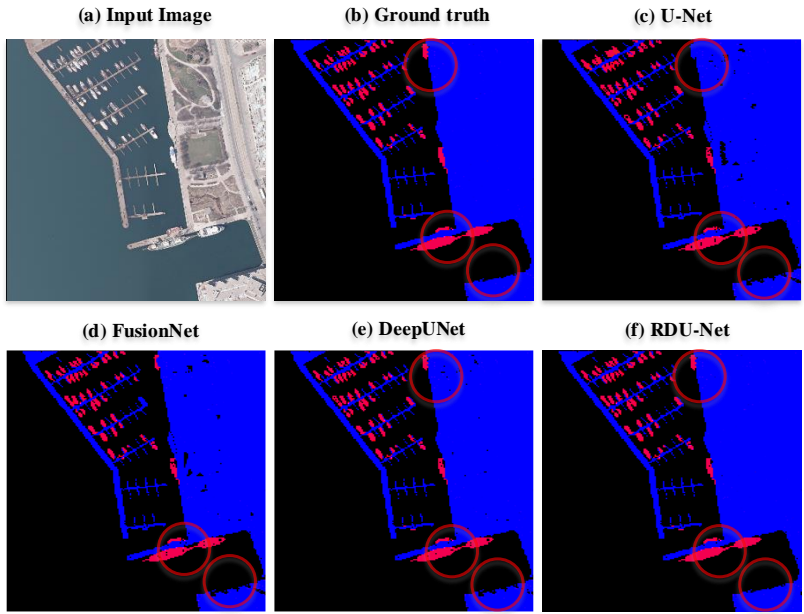

Fig. 8. The segmentation result of the high-resolution complex area by different models.

TABLE V
FIG. 7 DETAILS EVALUATION RESULTS

| Models | Precision | Recall | $F_1$ | Accuracy |
|---|---|---|---|---|
| U-Net | 0.9020 | 0.9002 | 0.8916 | 0.9183 |
| FusionNet | 0.9162 | 0.9099 | 0.9013 | 0.9174 |
| DeepUNet | 0.9498 | 0.9618 | 0.9429 | 0.9516 |
| RDU-Net | **0.9663** | **0.9689** | **0.9610** | **0.9727** |

TABLE VI
FIG. 8 DETAILS EVALUATION RESULTS

| Models | Precision | Recall | $F_1$ | Accuracy |
|---|---|---|---|---|
| U-Net | 0.8542 | 0.8427 | 0.8333 | 0.8926 |
| FusionNet | 0.8830 | 0.8744 | 0.8701 | 0.9038 |
| DeepUNet | 0.9271 | 0.9196 | 0.9109 | 0.9231 |
| RDU-Net | **0.9539** | **0.9578** | **0.9517** | **0.9693** |

These images have been used to check the performance of RDU-Net when having several tiny boundaries and very small objects. It is obvious in the experiments, RDU-Net precisely segmented the input images, especially at the sea-land boundaries. On the other hand, U-Net [9], FusionNet [15], and DeepUnet [5] failed to deal with the sensitive areas and there were several misjudgments on the tiny and complex areas.

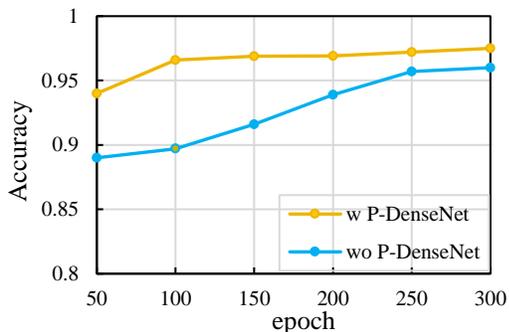

Fig. 9. The performance of RDU-Net with and without proposed DenseNet blocks on the whole dataset.

The performance of RDU-Net with and without the proposed DenseNet blocks is shown in Fig. 9. It is observed that the proposed model has significant improvement by adopting the proposed DenseNet blocks. Furthermore, while using the proposed DenseNet blocks, the model does not require many iterations to reach the highest rate, which is important in reducing the computation cost. We also present in Fig. 10 how the computation is distributed over 9 blocks in the original DenseNets and the proposed gradual DenseNets that shows the proposed model requires much less computational resources. It is notable that almost half of the computation cost belongs to the last two blocks which have high resolutions, and needs more time to construct them.

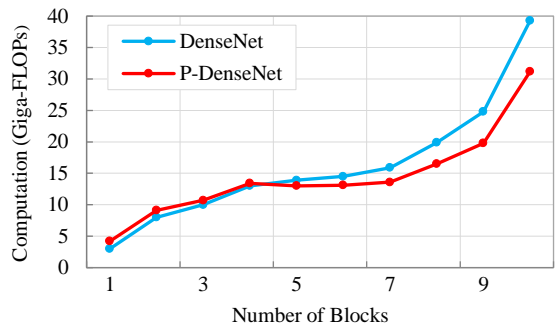

Fig. 10. The computation cost (in FLOPS) distribution over 9 blocks in DenseNet and proposed gradual DenseNet. Almost half of the computations are belong to the final two blocks due to the number of parameters and final results.

To analyze the effect of the proposed DenseNet, we use the evaluation metrics proposed in [24]. In Fig. 11, we present the performance of the proposed DenseNet. K represents the effects of each dense block onto the corresponding gate. The highest values belong to the blocks of indexes 1 to 4 in the downsampling path and 9 in the upsampling path. The lowest values are for the blocks of indexes 5, 6, 7 and 8 that are in the downsampling path which mostly performs the feature map dimension. This shows that, in such DenseNet blocks, the convolution layers are mostly used to increase dimension is more important than the DenseNet blocks itself.

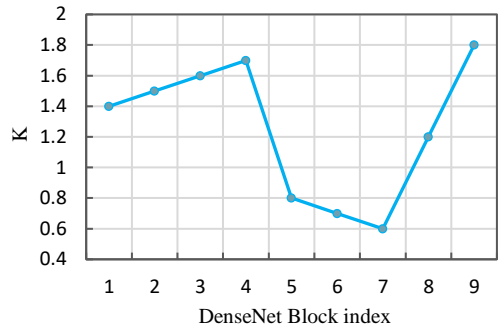

Fig. 11. Values for k according to ascending order of residual blocks. The first block, consisted of the first DenseNet block, has index 1, while the last block right before the Softmax layer has index 9.

TABLE VII
THE OVERALL TESTING EVALUATION RESULTS ON OUR DATASET FOR 300 EPOCHS.

| Models | Precision | Recall | $F_1$ | Accuracy |
|---|---|---|---|---|
| RF [39] | 0.6942 | 0.6994 | 0.6987 | 0.7061 |
| SVM [35] | 0.6431 | 0.6447 | 0.6450 | 0.6484 |
| U-Net [9] | 0.9438 | 0.9489 | 0.9347 | 0.9422 |
| SegNet [13] | 0.9413 | 0.9462 | 0.9336 | 0.9411 |
| ResNet [11] | 0.9411 | 0.9457 | 0.9366 | 0.9449 |
| Basaeed et al.[17] | 0.9592 | 0.9619 | 0.9562 | 0.9612 |
| FusionNet [15] | 0.9595 | 0.9621 | 0.9562 | 0.9613 |
| DenseNet [12] | 0.9598 | 0.9635 | 0.9572 | 0.9641 |
| Nogueira et al.[24] | 0.9635 | 0.9680 | 0.9597 | 0.9645 |
| DeepUNet [5] | 0.9642 | 0.9687 | 0.9603 | 0.9651 |
| RDU-Net | **0.9713** | **0.9706** | **0.9719** | **0.9739** |

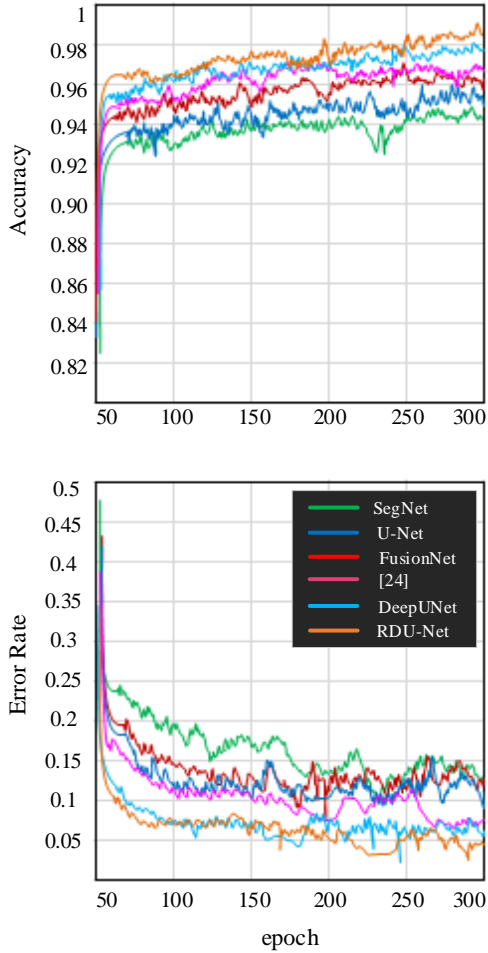

Fig. 12. The overall Accuracy and Error Rate assessment of 6 deep learning models on the whole dataset.

Fig. 12 shows the overall segmentation accuracy and error rates of RDU-Net in comparison with those of the other five deep learning models on the whole dataset. The accuracy results show the RDU-Net has near 98% segmentation accuracy and near 4% error rates, which show the superior performance of the proposed model against that of the other models. DeepUNet has the overall accuracy of 96.5% and the error rate of 7%. The Multi-Scale Segmentation model [24] has 96.3% accuracy and 9% error rate. The FusionNet, U-Net, and SegNet have the accuracy of 96%, 95% and 94% with the error rates of 9%, 12%, and 14% respectively. To assess the overall performance of the proposed RDU-Net, on the whole dataset, we compare its results with those of the eight deep neural network models (U-Net [9], SegNet [13], ResNet [11], FusionNet [15], Basaeed et al. [17], Nogueira et al. [24], DenseNet [12] and DeepUNet [5]) and two classical machine learning models (SVM and Random Forest). The results are illustrated in Table VII. The classical machine learning models do not have satisfactory performance, e.g. SVM precision result is 64.31% and its overall accuracy is 64.84%, while Random Forest is of 69.42% precision and 70.61% accuracy compared to SVM. On the other hand, to compare the results of the deep neural network models, firstly, RDU-Net has the best performance with 97.39% overall accuracy and 97.13% precision. Secondly, DeepUNet is of 96.51% accuracy which is better than FusionNet, DenseNet, ResNet, Basaeed et al. [17], Nogueira et al. [24], U-Net, and SegNet. Fig. 13 presents the computation cost and error rates of three deep learning models on the whole dataset. At the beginning, DeepUNet and RDU-Net have similar performances; gradually, RDU-Net outperforms the other models. The error rates difference between RDU-Net, DeepUNet and DenseNet are around 4% and 7% respectively.

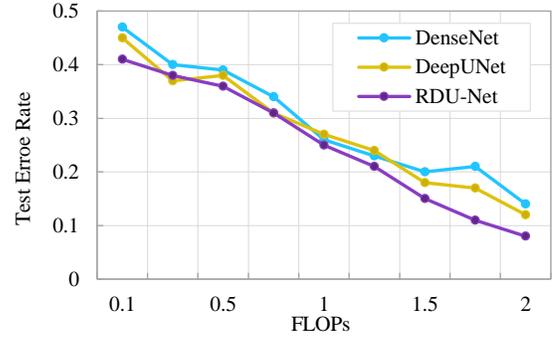

Fig. 13. On whole dataset, the Error Rate vs Computational Cost of 3 deep learning models.

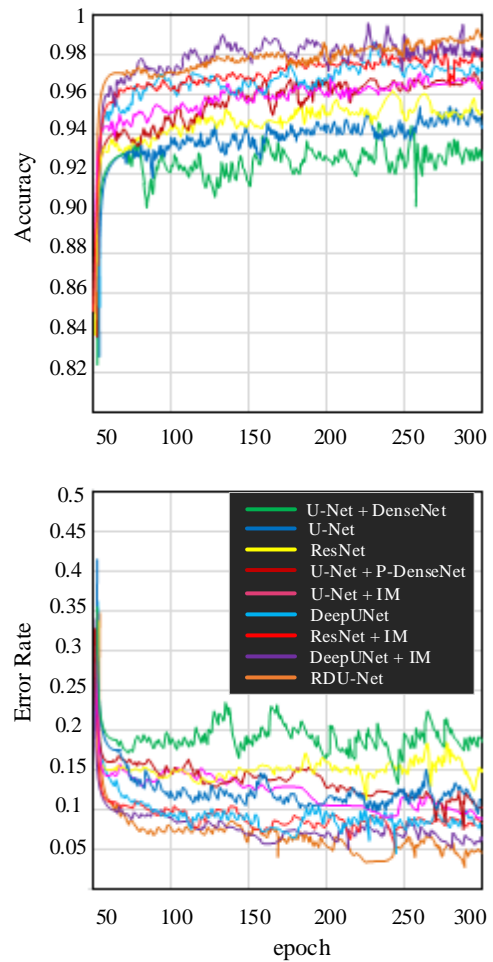

Fig. 14. The overall Accuracy and Error Rate assessment of several deep learning models with and without identity mapping (IM) and proposed DenseNet (P-DenseNet).

Fig. 14 shows the influence of identity mapping and the proposed DenseNet in different models. As the results show, adding the standard DenseNet to the U-Net decreases the performance. However, combining the proposed DenseNet and identity mapping with U-Net improves the results individually. Identity mapping improves the performance of ResNet [11] and DeepUNet [5]. It is worth to mention that by adding identity mapping to DeepUNet, its performance is very close to that of RDU-Net.

TABLE VIII
COMPARISON OF SEGMENTATION ERROR RATE ON MASSACHUSETTS ROADS AND SEA-LAND (SL) DATASET FOR 300 EPOCHS, GIGA FLOPS AND MILLION PARAMETERS.

| Models | FLOPs (G) | Params (M) | Massachusett dataset | SL dataset |
|---|---|---|---|---|
| ResNet [11] | 13.13 | 24.396 | 0.1688 | 0.1131 |
| U-Net [9] | 13.37 | 26.679 | 0.1679 | 0.1084 |
| SegNet [13] | 15.72 | 27.493 | 0.1584 | 0.1022 |
| DenseNet [12] | 40.69 | 35.475 | 0.1397 | 0.0941 |
| FusionNet [15] | 15.42 | 29.837 | 0.1451 | 0.0913 |
| Basaeed et al.[17] | 16.48 | 29.927 | 0.1421 | 0.926 |
| RU-Net [37] | 15.68 | 29.539 | 0.1424 | 0.0963 |
| Nogueira et al.[24] | 16.71 | 30.763 | 0.1383 | 0.0852 |
| DeepUNet [5] | 19.86 | 31.648 | 0.1353 | 0.0865 |
| RDU-Net | 15.44 | 18.617 | 0.1287 | 0.0748 |

Table VIII shows the results of RDU-Net and several state-of-the-art, efficiency of the models on the Massachusetts roads dataset [52] and sea-land dataset. From the results, we notice that RDU-Net requires approximately half parameters and FLOPs to achieve comparable accuracy to the original DenseNet. Somewhat surprisingly, our model even has better computation efficiency compared to DeepUNet [5] and RU-Net [37]. Furthermore, RDU-Net does not use depth-wise divisible convolutions, and just uses the simple convolutional filters with the sizes of 1×1 and 3×3. It is possible to use RDU-Net as a meta-architecture reported in [48] to obtain a more efficient network.

IV. CONCLUSION

In this paper, we have presented a novel deep neural network, Residual Dense U-Net (RDU-Net), for the purpose of sea-land remote sensing image segmentation. RDU-Net is based on the standard U-Net plus the proposed Densely Connected Residual Network blocks and several short-range and long identity mapping connections that construct a deep and appropriate network that has the ability to extract deep features and hierarchically re-use them to achieve accurate end-to-end image segmentation. We demonstrated the performance, low cost computation and flexibility of RDU-Net for the sea-land segmentation tasks. We compared RDU-Net against U-Net, FusionNet, DeepUnet and other models. The experimental results confirmed that RDU-Net outperformed the other state-of-the-art approaches in a number of quality metrics. In the future, we plan to build a more efficient RDU-Net to improve the segmentation accuracy not only on sea-land but also on the other segmentation tasks whilst reducing the overall computation costs.